\documentclass{aa}  
\usepackage{graphicx}
\usepackage{txfonts}

\usepackage{natbib}
\usepackage{hyperref}
\usepackage{xcolor}
\definecolor{DarkBlue}{rgb}{0.0,0.1,0.4}
\definecolor{Red}{rgb}{0.6,0.2,0.1}
\definecolor{DarkRed}{rgb}{0.5,0.0,0.5}
\definecolor{Blue}{rgb}{0.0,0.0,1.0}
\definecolor{Zelinkava}{rgb}{0.2,0.5,0.2}
\definecolor{Pink}{rgb}{0.0,0.7,0.7}
\definecolor{White}{rgb}{1.0,1.0,1.0}
\hypersetup{
    colorlinks=true,
    linkcolor=Blue,
    citecolor=Blue,
    filecolor=Blue,     
    urlcolor=Blue,
}
\newcommand\ext{\text{ext}}
\newcommand\rot{\text{rot}}
\newcommand\kms{\text{km}\,\text{s}^{-1}}
\newcommand{\de}{\text{d}}
\usepackage[normalem]{ulem}

\begin{document}

   \title{Intrinsic polarization of Wolf-Rayet stars due to
the rotational modulation of the stellar wind}

   \subtitle{}

   \author{S. Abdellaoui
          \inst{1}
          \and
          J.~Krti\v cka
          \inst{1}
          \and P.~Kurf\"urst\inst{2,1}}
   \institute{Department of Theoretical Physics and Astrophysics,
Faculty of Science,
           Masaryk University, Kotl\'a\v rsk\' a 2, 
Brno, Czech
           Republic,\\ \email{slah@physics.muni.cz}\and          
              Institute of Theoretical Physics, Charles University,
              V~Hole\v sovi\v ck\'ach 2, 
Praha 8, Czech Republic
             }

   \date{}

  \abstract
%
   {Fast rotating Wolf-Rayet stars are expected to be progenitors of long duration
   gamma-ray bursts. However, the observational test of this model is problematic. 
   Spectral lines of Wolf-Rayet stars originate in expanding stellar wind, therefore
   a reliable spectroscopical determination of their rotational velocities is difficult.
   Intrinsic polarization of Wolf-Rayet stars due to the rotational modulation
   of the stellar wind may provide an indirect way to determine the
   rotational velocities of these stars. However, detailed wind models are
   required for this purpose.
   }
   {We determine the intrinsic polarization of Wolf-Rayet stars from hydrodynamical
    wind models as a function of rotational velocity.}
   {We used 2.5D hydrodynamical simulations to calculate the structure of rotating
   winds of Wolf-Rayet stars. The simulations account for the deformation of the
   stellar surface due to rotation, gravity darkening, and nonradial forces.
   From the derived models, we calculated the intrinsic stellar polarization.
   The mass loss rate was scaled to take realistic wind densities of Wolf-Rayet stars  into account.}
   {The hydrodynamical wind models predict a prolate 
   wind structure, which leads
   to a relatively low level of polarization. 
   Even relatively large rotational velocities are allowed by observational constrains.
   The obtained wind 
   structure is similiar to that obtained previously for rotating optically thin winds.}
  %
   {Derived upper limits of rotational velocities of studied Wolf-Rayet stars are not in conflict with the model of long duration gamma-ray bursts.}
  %
%
%

   \maketitle
%

\section{Introduction}
Classical Wolf-Rayet (WR) stars are massive stars that lost their hydrogen envelope during 
their
evolution
\citep{1975MSRSL...9..193C,1979A&A....74...62C,2012A&A...540A.144S}.
The collapse of  fast rotating WR stars to a black hole could generate long duration gamma-ray
bursts \citep{1993ApJ...405..273W}.
In this context, \citet{2005A&A...442..587V} proposed that 
fast rotating
WR stars are progenitors of a  long duration
gamma-ray burst.
However, a direct observational test of this model is difficult due to the problematic
determination of rotational velocities of Wolf-Rayet stars. An estimation
of rotational velocities from polarized light may provide an indirect way to test the 
nature of gamma-ray burst progenitors.
Winds of WR stars are ionized, consequently the light scattering on free
electrons is an important source of opacity in the envelopes of these stars. As
a result of light scattering on free electrons, an initially unpolarized 
beam coming 
from the stellar photosphere becomes linearly polarized. After
scattering into the direction making an angle of $\Theta$ with respect to the
incident beam, the ratio of intensities in the directions perpendicular and
parallel to the plane of scattering (given by the directions of incident and
scattered light) is $1:\cos^2\Theta$ \citep{1950ratr.book.....C}. However, in
spherically symmetric envelopes, the contribution of light polarized in different
directions cancels out resulting in zero net polarization.

Most stars are known to be more or less spherical,  but the rapid rotation could break the sphericity and lead to an axisymmetric density structure.
If electron scattering takes place in an axisymmetric envelope, the intrinsic polarization occurs. 
From the measurement of \citet{1970ApJ...160.1083S}, few WR stars have been found to be 
intrinsically polarized. \citet{1979ApJ...231L.141M} discussed the dependence 
of the continuum polarization on wavelength in WR  stars and discovered the decrease in polarization across the emission features.

To interpret the observed  polarization, \citet{1977A&A....57..141B}  developed an analytical 
method to compute the polarization due to optically thin electron scattering in axisymmetric envelopes.
They assumed that the continuum polarization mainly depends on the inclination angle toward the 
line of sight, the geometry, and the optical depth.  \citet{1987ApJ...317..290C}  
rederived the polarization expression of 
\citeauthor{1977A&A....57..141B} and added the depolarization correction factor of finite stellar size. 
\citet{2000A&A...356..619B} showed that the depolarization factor has a  
negligible effect on the value 
for the polarization.

In the observational context \citep{1994Ap&SS.221..347S},
there are several methods to separate the interstellar polarization and the 
intrinsic polarization,  each being based on different assumptions. 
\citet{2018MNRAS.479.4535S}  applied Serkowski's law to fit the interstellar polarization of WR93b and WR102.
They conclude that these two stars are not intrinsically polarized, hence no line effect 
could be found,  resulting from the dilution of continuum polarization by unpolarized lineflux. 
They placed the upper limit on the rotational velocity, which is lower than the value determined by
\citet{2012A&A...540A.144S} from the shape of emission line
profiles, where WR93b and WR102 are among a small group of WR stars with round emission line profiles, possibly implying fast rotation. High rotational velocities are required to form a collapsar, which is likely a source of long duration gamma-ray bursts \citep{1999ApJ...516..788W}.

In this work, we discuss the limits of the rotational velocity of WR stars 
from polarimetry, in addition to the dependence of  polarization on the inclination angle 
and rotational velocity.
Our study is based on hydrodynamic stellar  wind models of the selected WR stars WR93b and WR102 that include the effect of rotation.
The obtained density was used to calculate the residual polarization. In 
Section \ref{comp}, we describe the method by which the polarization was calculated. 
Sections \ref{result}-\ref{discus} provide the discussion of the results with some comparison 
to the results of \citet{2018MNRAS.479.4535S}.
In the last section, we provide the conclusions and some suggestions  that might prove fruitful in the future. 
%
 
\section{Computational method} \label{comp}
\subsection{Hydrodynamic model}
Stellar wind of hot stars is described by coupling radiative force with 
hydrodynamic model equations. 
The radiative force is based on the Sobolev approximation for stellar wind
\citep{1960mes..book.....S,1974MNRAS.169..279C,1995ApJ...440..308C}. The
hydrodynamic model equations describing the wind expressed in spherical geometry
are as follows in the axisymmetric case
\citep{1994ApJ...424..887O}: 

\begin{align}
\label{eq1}
\frac{\partial\rho}{\partial t}+\frac{1}{r^{2}}\frac{\partial(r^{2}\rho 
\varv_{r})}{\partial r}+\frac{1}{r\sin\theta}\frac{\partial(\rho 
\varv_{\theta}\sin\theta)}{\partial\theta} & =0,\\ 
\frac{\partial \varv_{r}}{\partial t}+\varv_{r}\frac{\partial \varv_{r}}{\partial r}+\frac{\varv_{\theta}}{r}\frac{\partial \varv_{r}}{\partial\theta}-\frac{
\varv_\theta^{2}+
\varv_{\phi}^{2}}{r}  &= -\frac{1}{\rho}\frac{\partial P}{\partial r}+  g_{r}^{\ext} ,\\
\frac{\partial \varv_{\theta}}{\partial t}+
\varv_{r}\frac{\partial \varv_{\theta}}{\partial r}+\frac{\varv_{\theta}}{r}\frac{\partial 
\varv_{\theta}}{\partial\theta} - \cot\theta\frac{
\varv_{\phi}^{2}}{r}+\frac{
\varv_{r}\varv_{\theta}}{r}&=-\frac{1}{r\rho}\frac{\partial P}{\partial 
\theta
}+ g_{\theta}^{\ext},\\
\frac{\partial 
\varv_{\phi}}{\partial t}+
\varv_{r}\frac{\partial 
\varv_{\phi}}{\partial r}+\frac{
\varv_{\theta}}{r}\frac{\partial \varv_{\phi}}{\partial\theta} +\cot\theta\frac{
\varv_\theta\varv_\phi}{r}+\frac{\varv_{r}\varv_{\phi}}{r} &=g_{\phi}^{\ext},\\
P&=  c_\text{s}^2\rho \label{eq2},
\end{align}
where $\rho$ is the density of the fluid, $\varv_r$, $ \varv_\theta$, and $\varv_\phi$ 
are the velocity components in every direction, $c_\text{s}$ is the sound speed, and $P$ is the 
gas pressure.
For a star with a radius $R_\star$, mass  $M_\star$, luminosity  $L_\star$, and temperature $T_\star$,  the external force $\vec{g}^{\ext}$ is  the vector sum of the line radiative force  $\vec{g}_{\text{line}}$ and the effective gravity  $\vec{g}_{\text{eff}}$.

The line force $\vec{g}_{\text{line}}$ at the position $\vec{r}$  was obtained by integrating numerically the stellar 
intensity $I$ over the solid angle $\Omega_\star$ weighted by the velocity gradient \citep{1995ApJ...440..308C}

\begin{equation}
 \vec{g}_\text{line}(\vec{r})=\frac{k \sigma_\text{T}^{1-\alpha}}{
 \rho(\vec{r})^{\alpha}c\varv_{\text{th}}^\alpha}\int_{\Omega_{\star}}(\vec{n}\cdot\mathbf{\nabla}[\vec{n}\cdot
 \vec{\varv}(\vec{r})])^{\alpha}\vec{n}\,I(\vec{n},\vec{r})\,\de\Omega.
\label{gline}
\end{equation}

Here, $\alpha$ and $k$ are the  \citet[hereafter CAK]{1975ApJ...195..157C} and \citet{1982ApJ...259..282A}
parameters for the radiative force, respectively,  $\sigma_\text{T}$ is the cross section due to Thomson scattering on free electrons, $\varv_{
\text{th}}$ is the thermal velocity,  and $c$ is the speed of light. 
We note that the line force also depends on the dilution factor due to ionization which has been removed here as  the $\delta$ parameter can be put equal to zero (see Table~\ref{cakparam}).

To solve the hydrodynamic model equations  \eqref{eq1} -- \eqref{eq2}, we  used the VH-1 \citep{1990ApJ...356..591B} piecewise parabolic method (PPM) code  coupled with the subroutine developed by \citet{1996ApJ...472L.115O} to  determine the radiative force. 
The well-known PPM algorithm was developed by \citet{1984JCoPh..54..174C}, and it is the third-order  finite difference scheme. The VH-1 code solves the hydrodynamic equations in 1D, 2D, and 3D planar,  cylindrical, and spherical geometry. For our problem, we used  2.5D spherical geometry, with three velocity components ($\varv
_r$, $\varv
_\theta$, $\varv
_\phi$), where the flow is specified along the radial and 
colatitudinal coordinate components ($r$ and $\theta$). 
The grid is discretized into $n_i=320$ zones in radius, from stellar 
surface $r_{\min}= R_\star$ to $r_{\max} \approx 10R_\star$.
The latitudinal direction is subdivided into $n_j=100$ equidistant zones 
from the north pole ($\theta=0$) to the south pole ($\theta=\pi$). 

The most important factor for hydrodynamic simulation of stellar winds is to 
define the lower boundary condition in the radial direction to introduce the 
oblate stellar surface due to rotation. \citet{1994ApJ...424..887O} proposed 
a specified  stair-casing boundary condition to solve this difficulty
We have chosen the same configuration to formulate our problem. To assure a 
subsonic inflow of matter at the base of the wind, we set the base wind 
density to a constant value proportional to the mass loss rate $\dot{M}$,
\begin{equation}
\rho(r=R_\star)\propto \frac{\dot{M}}{4\pi R_\star^2c_\text{s}},
\end{equation}
whereas for the velocity component, we assume a subsonic outflow for polar velocity and rigid body rotation for the azimuthal velocity.  To control the oblateness of the stellar surface, the radius was set as a function of colatitude $\theta$ and the ratio of rotational velocity and the critical velocity $\omega$
\citep{1966ApJ...146..152C, 1994ApJ...424..887O}
\begin{equation}\label{cubic}
    \frac{R_\star(\theta)}{R_p}=\frac{3}{\omega\sin\theta}\cos\left(\frac{\pi+\arccos(\omega\sin\theta)}{3} \right)
,
\end{equation}
where $R_\star(\theta)$ is the surface radius
and $R_p$ is the polar radius.
For the upper boundary, we set an outflow  condition.
Because of the symmetry, we set a reflecting boundary conditions (default from VH-1 code) in the latitudinal direction.

\subsection{CAK  parameter optimization}

In order to make the numerical simulation of the two WR stars, we have to know the CAK 
parameters of the line force. To our knowledge, there has not been a previous calculation of line force parameters for these two stars. 
Therefore, to get the value of $\alpha$ and $k$, we required the terminal velocity and mass loss rate from observations. We varied the 
line force parameters to fit the observational values. We adopted the stellar parameters from 
\citet{2015A&A...581A.110T} and \citet[see Table~\ref{starparam}]{2018MNRAS.479.4535S}. 
The selected values for CAK parameters are  given in Table~\ref{cakparam}. The critical velocity is $1734\,\kms$ for WR93b and $2286\,\kms$ for WR102 \citep{2018MNRAS.479.4535S}. 

The mass loss rate corresponding to the selected CAK parameters is $5.8 \times 10^{-7} \,M_\odot\,\text{yr}^{-1}$
for WR93b and $4.7 \times 10^{-7}\,M_\odot\,\text{yr}^{-1}$
for WR102, which is lower than what was derived from observations. Therefore, for the calculation of the polarization, we multiplied the density by the ratio of the observed mass loss given in Table~\ref{starparam} to the numerical mass loss from the hydrodynamic simulation.
We multiplied the final density by a factor of $17$ for WR93b and $24$ for WR102.
We could increase the mass loss rate to 
the observational values (e.g., including the $\delta$ parameter),
but running the model with rotation leads
to a nonphysical structure of the density and the velocity, as we could get higher velocity at the polar regions and some numerical instability occurred inside the computational domain.
The obtained terminal velocity is about $6000\,\kms$ for WR93b and around $8000\,\kms$ for WR102.
\begin{table*}
\caption{Adopted stellar and wind parameters.}
\label{starparam}
\begin{tabular}{ccccccc}
\hline 
\hline 
WR & $\log (L_\star/L_{\odot})$ & $T_{\star}$ & $R_{\star}$ & $M_\star$ &$\varv_{\infty}$ & $\log(\dot{M}/1\,M_{\odot}\,\text{yr}^{-1}) $ \\
&&    (kK) & $(R_{\odot})$& ($M_\odot$) & ($\text{km}\,\text{s}^{-1}$) &\tabularnewline
\hline 
93b & 5.30 & 160 & 0.58 & 
7.1 & 5000 & -5 \tabularnewline
102 & 5.45 & 210 & 0.39 & 
7.0 & 5000 & -4.92\tabularnewline
\hline 
\end{tabular}
\end{table*}

\begin{figure}
\includegraphics[scale=0.6]{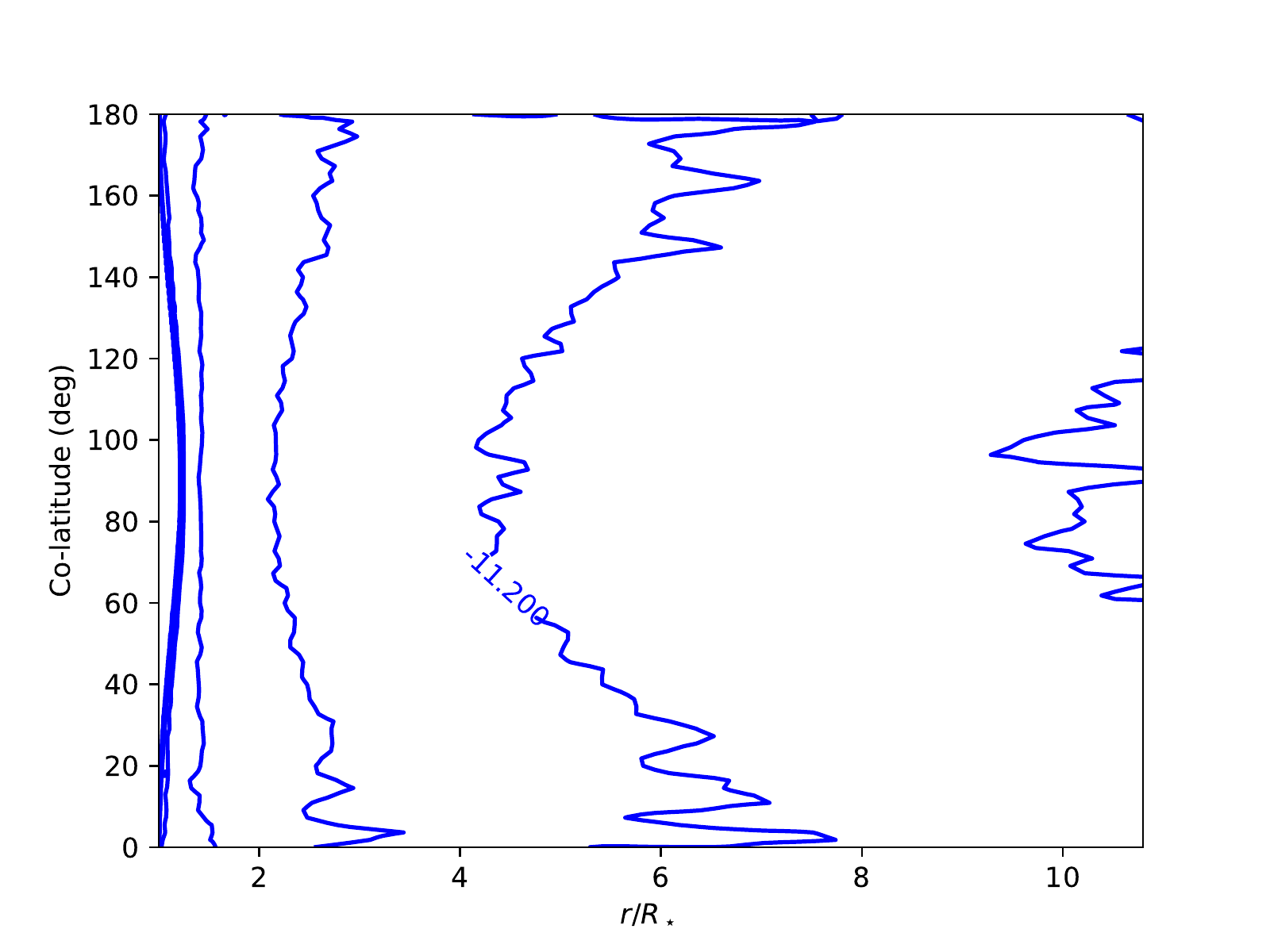} 
\caption{Density (in g$\cdot$cm$^{-3}$)
contours of stellar wind of the star WR93b versus radius $r$ and colatitude $\theta$, in log scale spaced by 0.8 dex, denoted contour correspond to $\log(\rho)=-11.2$, for rotation $V_{\rot}=1100\,\kms$, with nonradial forces and gravity darkening.}%
\label{dens102}
\end{figure}

We ran the  simulations for different rotational velocities. 
To check the influence of individual assumptions on the intrinsic polarization, we
calculated several sets of wind models with an increasing level of complexity.
First, we included radial force only leading to a compression effect as discussed by \citet{1993ApJ...409..429B}. Within these assumptions, an equatorial density enhancement forms, as found by \citet{1994ApJ...424..887O}. 
Second, we added the nonradial forces,  which inhibit the formation of the disk \citep{1996ApJ...472L.115O}. In the most complex model, we also included the gravity darkening. 
Figure \ref{dens102} shows the density contours as a function of the radius and colatitude for $V_{\text{rot}
}=900\,\kms$; the results are similar to those found by  \cite{1996ApJ...472L.115O} and show a reduced density at the equator due to the gravity darkening and the effect of nonradial forces.
Our aim is not to investigate  the equatorial disk formation, but rather we want to make use of the hydrodynamic simulation in order to obtain the resulting polarization. In the following section, we  show 
how the continuum polarization can be calculated from the wind density $\rho$ derived from the hydrodynamic model.
\begin{table} 
\caption{Adopted CAK parameters.}\label{cakparam}
\begin{tabular}{cccc}
\hline 
\hline
 & $\alpha$ & $k$ & $\delta$\tabularnewline
\hline 
WR93b & 0.52 & 0.61 & 0\tabularnewline
WR102 & 0.52 & 0.61 & 0\tabularnewline
\hline 
\end{tabular}
\end{table}

\subsection{Polarization model}

\begin{figure}
\includegraphics[scale=0.85]{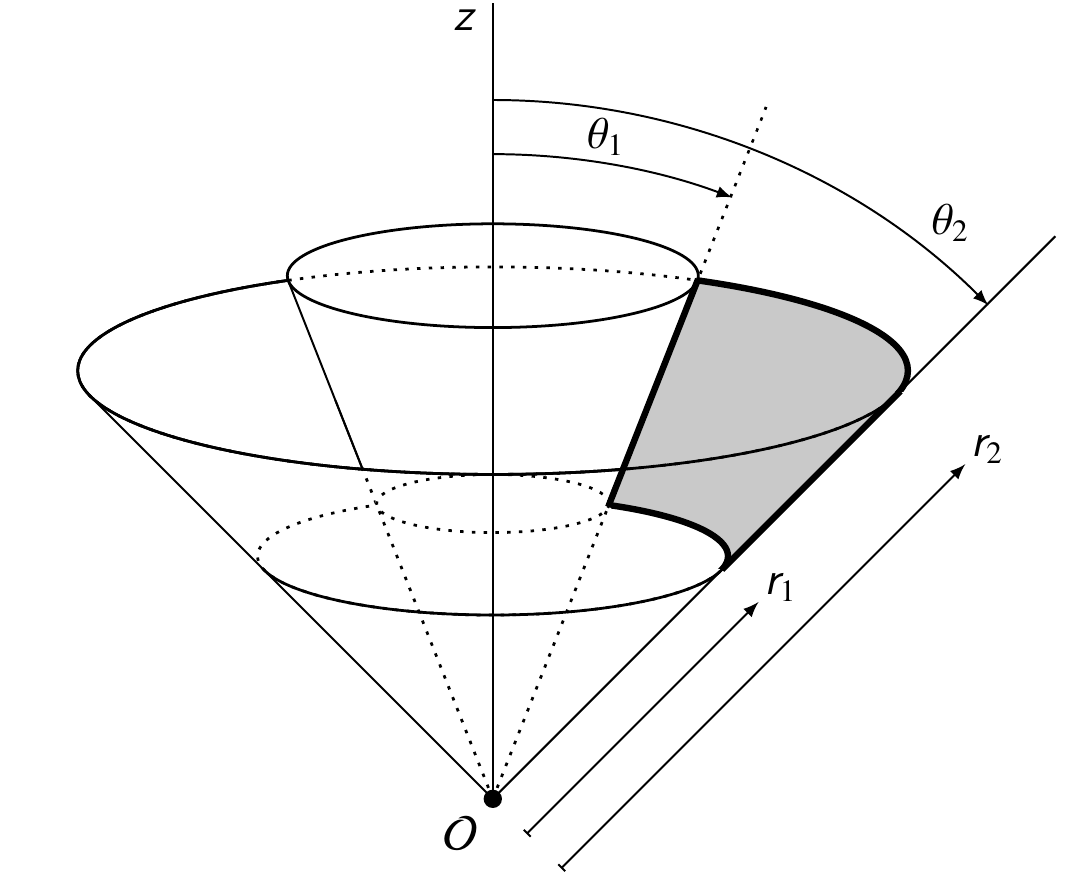} 
\caption{
Geometry of the plane of integration in spherical geometry ($r$,$\theta$,$\phi$) adopted from \citet{2000A&A...356..619B}.}
\label{geom}
\end{figure}

We intend to model the continuum polarization due to electron scattering, as we expect that the polarization will vanish in the emission lines.
Polarization from radiation scattering depends on the integral of electron density and the volume of scattering. The analytic expression of the
polarization was developed by \citet{1977A&A....57..141B}, which  depends on the inclination angle, optical  depth, and geometry factor, as follows:
\begin{equation}
P_R \approx \bar{\tau}(1-3\gamma)\sin^{2}i,
\label{polar}
\end{equation}
where $\bar{\tau}$ is the  averaged Thomson scattering optical depth defined as the integral of the number density over  the  radius and the colatitudinal angle $\theta$ (as shown in Fig. \ref{geom}, see also \citealt{1978A&A....68..415B}),
\begin{equation}
\bar{\tau} = \frac{3}{16}
\sigma_\text{T}
\int_{r_1}^{r_2}\int_{\mu_1}^{\mu_2} n_\text{e}(r,\mu)
\,\de r\,\de\mu.
\label{Thomson}
\end{equation}
 
The factor $\gamma$ in Eq.~\eqref{polar} is given by
\begin{equation}%
\gamma = \frac{\int_{r_1}^{r_2}\int_{\mu_1}^{\mu_2}n_\text{e}(r,\mu)
\,
\mu^2
\,\de r\,\de\mu
}{\int_{r_1}^{r_2}\int_{\mu_1}^{\mu_2}n_\text{e}(r,\mu)
\,\de r\,\de\mu},
\end{equation}
where $\mu=\cos\theta$, 
$i$ is the inclination angle, and $n_\text{e}$ is the electron number density, which depends on the mass density $\rho$;
\begin{equation}
n_e= \frac{\rho}{\mu_\text{e} m_\text{H}},
\end{equation} 
where 
$m_\text{H}$
is the  hydrogen atom  mass and $\mu_\text{e}$ is the mean molecular weight per free electron \citep[cf. also the calculation of polarization signatures in][]{2020A&A...642A.214K}.
To take the depolarization effect of finite stellar geometry into account, \citet{1987ApJ...317..290C}  introduced the correction factor  $D(r)=\sqrt{1-R^2_\star/r^2}$, and the polarization relation becomes the following:
\begin{equation}
\label{polarcor}
        P _R= \frac{3}{16} \sigma_\text{T}\sin^2 i \int_{r_1}^{r_2}\int_{\mu_1}^{\mu_2}n_e(r,\mu)(1-3\mu^2)D(r) \, \de r \, \de\mu. 
    \end{equation}
Stellar winds of WR stars are optically thick even in the vicinity of the sonic point. However, Eq.~\eqref{polarcor} accounts for just the polarization due to optically thin scattering. Therefore, we selected the lower limit of integration in Eq.~\eqref{polarcor} as $r_1=1.2R_\star$ to exclude the optically thick part of the wind. The upper limit corresponds to the outer boundary of numerical simulations, $r_2=10R_\star$. 
\subsection{Polarization due to surface temperature variation}
The shape of the star changes due to the effect of rotation becomes oblate by 
the effect of centrifugal forces, and the emergent flux experiences a redistribution so that the polar regions become bright, and the equator becomes dark, what is known as the effect of gravity darkening.
These effects can be specified by the known Roche and von Zeipel models for early type stars \citep{1968ApJ...151.1051H,1995ApJ...440..308C} and they affect the radiative intensity in the wind. The anisotropic intensity introduces an additional source of polarization.

The von Zeipel theorem \citep{1963ApJ...138.1134C,1968ApJ...151.1051H,1995ApJ...440..308C} shows that the local flux $F(\theta)$ 
on the surface of a rotating star is proportional to local effective gravity $g(\theta)$
\begin{equation}
    F(\theta)=\sigma_BT_\text{eff}^4(\theta)\propto g_\text{eff}(\theta),
\end{equation}
where $\sigma_B$ is the Boltzmann's constant, $T_\text{eff}(\theta)$ is the 
effective temperature depending on
colatitude, and the magnitude of surface gravity is expressed as
\begin{equation}
    g_\text{eff}(\theta) = \frac{GM_\star}{R_p^2}\left[\left(-\frac{1}{x^2}+\frac{8}{27}\omega^2x\sin^2\theta\right)^2+\left(\frac{8}{27}\omega^2x\sin\theta\cos\theta\right)^2 \right]^\frac{1}{2}
,\end{equation}
where $GM_\star/R_p^2$ is the polar gravity term at the critical velocity. 
von Zeipel theorem does not provide the constant of proportionality and many authors  \citep{1963ApJ...138.1134C, 1995ApJ...440..308C} fitted this constant as a function of $\omega$; we assume this constant to be unity for simplicity.

If we assign the temperature to a point in the wind, the specific intensity $I$ can be computed using 
\begin{equation}
    I_0(\vec r,\vec n) = B_\nu(T_\text{eff}(\theta))
,\end{equation}
where $B_\nu$ is the Planck function of a black body.

From the work of 
\citet[Eq.~(20)]{1977A&A....57..141B},
we have the intrinsic polarization 
\begin{equation}
    P_R\approx\frac{I_1}{I_0}
%
P_1,
\end{equation}
where $I_1 \ll I_0$ and $P_1$ is the polarization of the scattered intensity $I_1$
\begin{equation}
    P_1=\frac{3\pi\sigma_T\sin^2i}{16}\frac{\int_{r_1}^{r_2}\int_{\mu_1}^{\mu_2}n_\text{e}(r)B_\nu
    (T_\text{eff}
    (\theta))[1-3\mu^2]\de r\de \mu}{I_1}.
\end{equation}

Replacing $P_1$ and $I_0$ and by their expressions, we get
\begin{equation}\label{tempol}
    P_R=\frac{3\pi\sigma_T\sin^2i}{16}\frac{\int_{r_1}^{r_2}\int_{-1}^{1}n_\text{e}(r)(1-3\mu^2)B_\nu(
    T_\text{eff}
    (\theta))\de r\de \mu}{B_\nu(T_\star)}.
\end{equation}
It is possible to add the depolarization $D(r)$ effect due to finite stellar size as previously in Eq.~\eqref{polarcor}.

\subsection{Numerical integration}

In order to numerically evaluate the  integrals, we have used the trapezoidal rule as implemented in the Numpy  library \citep{harris2020array}. Moreover, we used Pandas for data analysis  \citep{mckinney2011pandas} and Matplotlib for plotting \citep{Hunter:2007}.

Calculations were carried out for various values of rotational velocities.
The results are shown in  Figs.~\ref{fig:rad93} -- \ref{polwr102}, where the polarization  $P_R$ is plotted as a function of the inclination angle $i$ for different  fractions of the critical velocity. 

\section{Results}\label{result} 

\citet{2018MNRAS.479.4535S} applied the simplified model of  wind-compressed zones
by \citet{1996ApJ...459..671I} to investigate the dependence of the linear  polarization on the stellar rotation. In our  model, we included gravity darkening and oblateness of the star due to rotation  effects; the model is based on the work  
of \citet{1995ApJ...440..308C} and \citet{1996ApJ...472L.115O}. 
 
\begin{figure}
    \centering
    \includegraphics[scale=0.6]{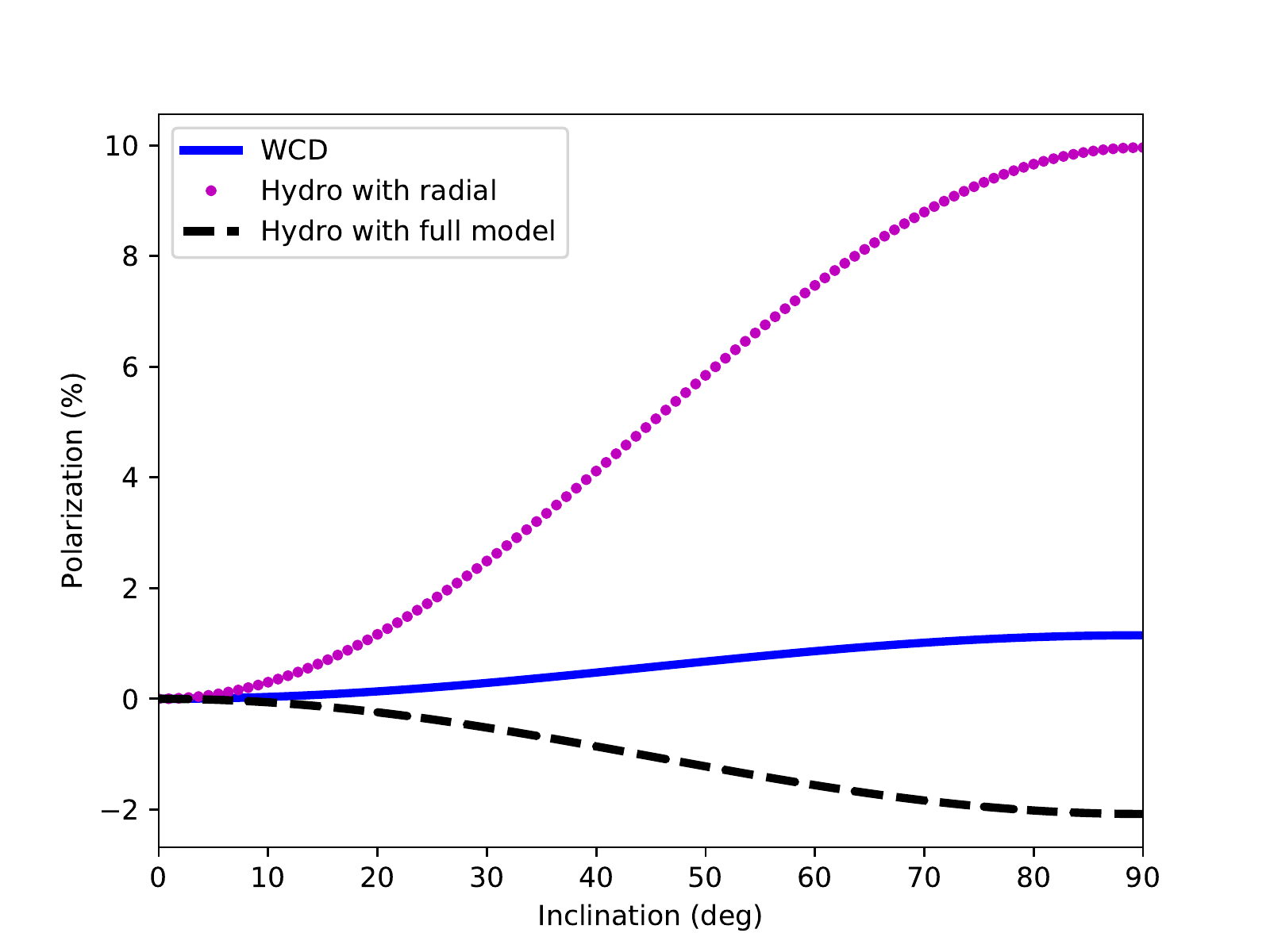}
    \caption{Percentage polarization of WR93b 
    from the model with
    radial force 
    only
    (hydro with radial), nonradial forces, and gravity darkening (hydro with full model), 
    compared to
    the model of the \citet{1996ApJ...459..671I} 
    wind compressed disk (WCD) for $V_\text{rot}=900\,\kms$.}
    \label{fig:rad93}
\end{figure}

\begin{figure}
    \centering
    \includegraphics[scale=0.6]{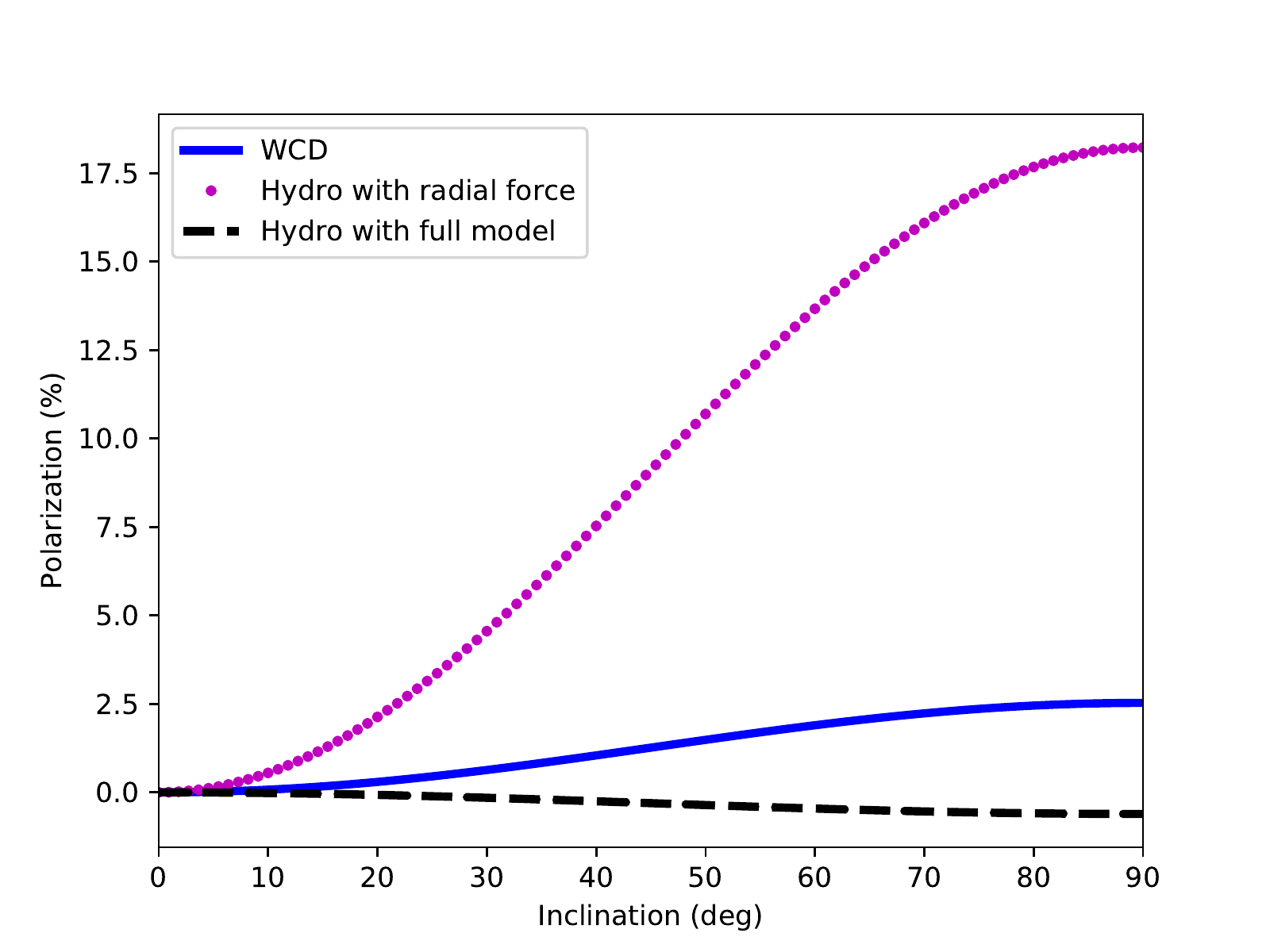}
    \caption{
    As in Fig.~\ref{fig:rad93}, but for WR102.}
    \label{fig:rad102}
\end{figure}

\begin{figure}
\centering
\includegraphics[scale=0.6]{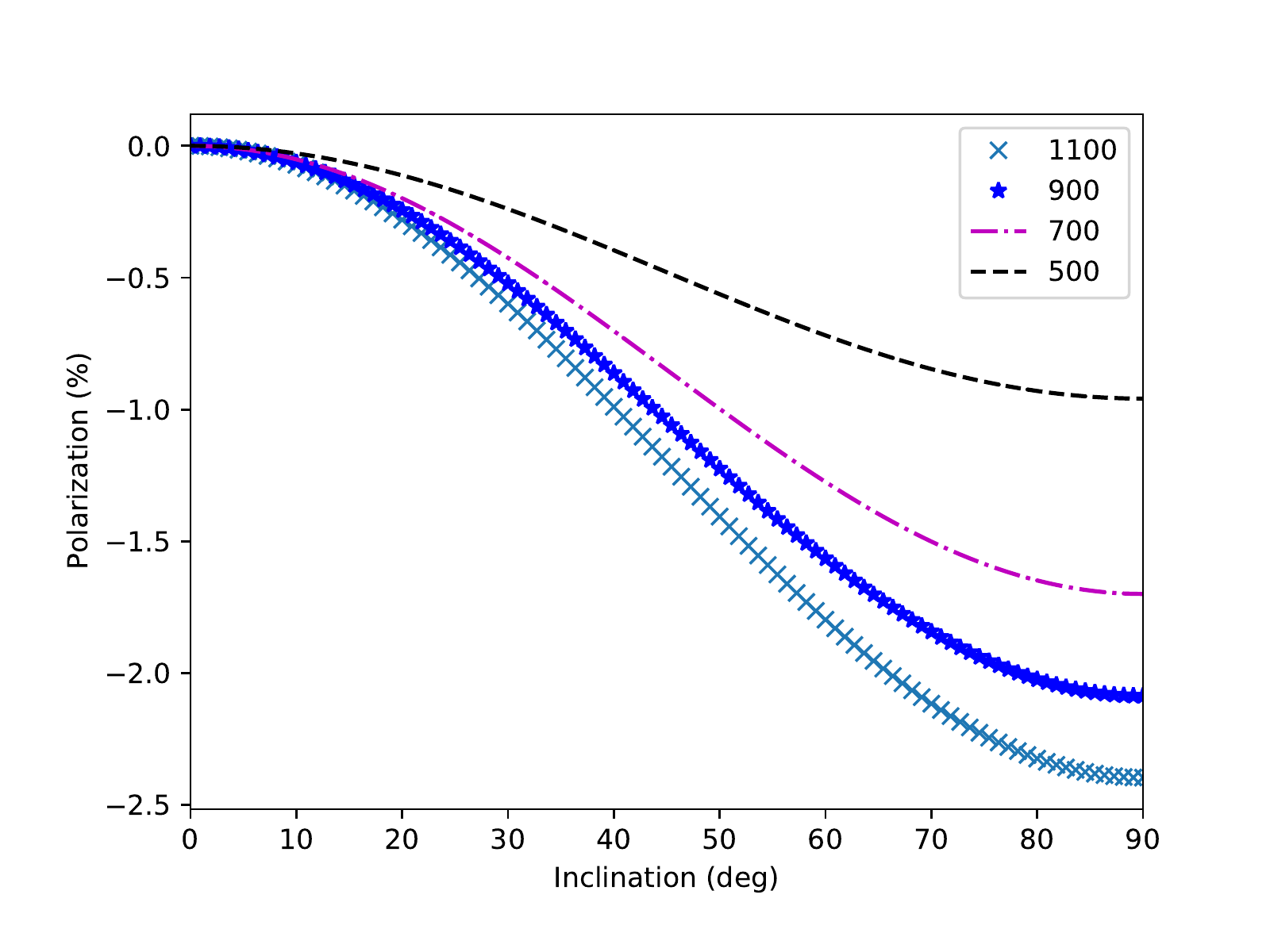} 
\caption{Polarization of WR93b as a function of the inclination for different rotational
velocities (denoted in the graph in units of $\kms$) determined from the models
including nonradial forces and gravity darkening.
}
\label{polwr93}
\end{figure}

\begin{figure}
\centering
\includegraphics[scale=0.6]{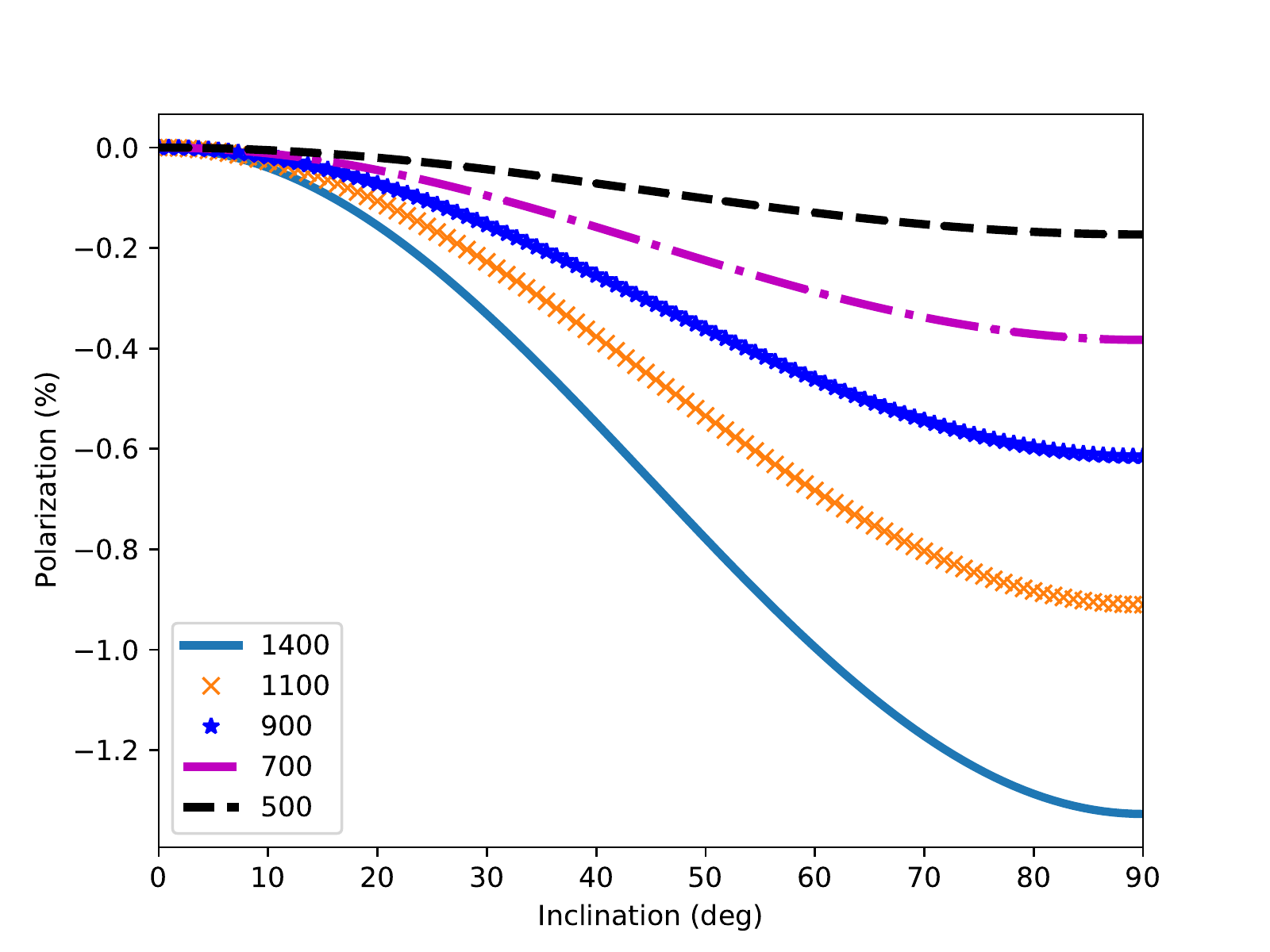} 
\caption{
As in Fig.~\ref{polwr93}, but for WR102.}
\label{polwr102}
\end{figure}

Figures \ref{fig:rad93}-- \ref{polwr102} show  an increase in absolute values of the polarization by
increasing the rotational velocity and the inclination angle. For higher rotational velocities, the departure from spherical symmetry becomes stronger, which leads to higher linear polarization.
The inclination is another crucial parameter to observe the intrinsic polarization.
When the inclination tends to zero, the
system is viewed in the direction  from the pole, the distribution of matter
becomes symmetric with respect to the viewing direction, and the polarization drops to zero. 
The polarization  reaches its maximum when the system is viewed equator on and when the angle coincides with the scattering plane.
The signature of polarization is given by the geometry. The signature is positive for oblate disk-like geometry, while it becomes negative for prolate jet-like geometry.

The upper limit for rotation velocities was set to $19\%$ and $10\%$ (of the critical velocity)
for both stars in the work of \citet{2018MNRAS.479.4535S}. 
This was determined from the observed value of the polarization, which is less than  0.077\%  for WR93 and less than  0.057\%  for  WR102. 
For a comparison with their results, we additionally employed the analytical model by
\citet{1996ApJ...459..671I}
to calculate the density structure of the wind. We calculated two sets of hydrodynamical models, one where the density was numerically computed including the radial force only, which corresponds to the analytical model. The second set of more realistic models includes the gravity darkening and nonradial components of the radiative force.

Due to a nonphysical structure which appears for low rotational velocities, we  chose to calculate the polarization for  the two stars with a higher rotational velocity than the ones  employed by \citet{2018MNRAS.479.4535S}, as seen in  Figs.~\ref{fig:rad93} -- \ref{fig:rad102}.  Figure~\ref{fig:rad93} shows  the intrinsic polarization for WR93b, where the values obtained from the hydrodynamic simulation  with only radial forces of about $10\%$ are higher than for the analytic  model viewed equator on, while for the model with gravity darkening and nonradial forces, the polarization is $2\%$ in absolute value.  
Figure~\ref{fig:rad102} illustrates the  variation of the polarization with inclination for WR102 star, the polarization from the hydrodynamic model is $18\%$, which is
higher than from
the analytical model ($2.52\%$), while from nonradial forces and gravity darkening it is $0.61\%$ in absolute value. 
In general, the polarization is positive for oblate density distribution and this shows  that the plane of polarization is parallel to the symmetry axis.
The polarization is negative when obtained from the model with gravity darkening, and this indicates that there is  a higher density of material at the polar region than at the equatorial region,  and the plane of polarization is perpendicular to the symmetry axis.

For the hydrodynamic model with the radial force only, we chose to integrate from the radius $r_1=1.2R_\star$; the reason is that the oblateness of the stellar geometry,  due to rotation and lower boundary conditions, rises the polarization degree and gives 
overestimated values.
\citet{1993A&A...271..492W} calculated the polarization for a Be star and they developed a relationship to get an estimate of the degree of the polarization. Applying equation (39),
\begin{equation}
    P_R=\frac{3\sigma_\text{T}}{32\pi}\frac{\dot{M}}{2\pi m_p R_\star \varv_{\infty}};\end{equation}
from their paper to the stellar parameters,  we see that  $P_R\approx 8.6\%$ for WR93b and $P_R\approx 18\%$ for WR102, which shows a good agreement with the hydrodynamic model. 


Figures~\ref{polwr93}--\ref{polwr102} show the 
variations of the polarization for the stars WR93 and WR102, respectively, and for different rotational velocities
by adding nonradial forces and the gravity darkening.
For the reason stated above, we chose to integrate from $r_1=1.2R_\star$.
The polarization is negative, which shows a prolate structure of the geometry,  and for low rotational velocities, the intrinsic polarization tends to 
$-0.95\%$
 for WR93b 
 and to $-0.17\%$ for WR102. Increasing the rotational velocity decreases the net polarization; for a rotation about $60\%$ of the critical velocity, we get $P_R=-2.4\%$ and $P_R=-1.32\%$ for WR93b and WR102, respectively.
 
 To obtain the upper limits for rotational velocity, we carried out a logarithmic least square fitting for the model with nonradial forces and gravity darkening. 
 Table~\ref{fitting}  shows a comparison of upper limits of rotational velocity, at the inclination angle
 $90^\circ$ (edge-on), from our model and the model of \citet{2018MNRAS.479.4535S}.
 
 Calculating the angular momentum $j$ from the obtained rotational velocity, and comparing this with the threshold  $j\geq3\times10^{16}\,\text{cm}^2\,\text{s}^{-1}$
 of the collapsar derived by \citet{1999ApJ...524..262M} in Table~\ref{fitting}, the lower limits of angular momentum exceed the threshold of \citet{1999ApJ...524..262M}. 
 Therefore, the studied WR stars may be progenitors of a long gamma-ray burst.
 This agrees with the results of \citet{2012A&A...547A..83G}, who determined the rotational velocities from photometry, and with the original results of \citet{2018MNRAS.479.4535S}.
\begin{table*}
    \caption{Comparison between the fitted
    maximum observationally allowed
    rotational velocity (in $\kms$) in this model and the model of
    \citet{2018MNRAS.479.4535S}.
    The last column gives the maximum specific angular momentum determined from the maximum allowed rotational velocity.}
    \label{fitting}
\begin{tabular}{ccccc}
\hline
\hline 
WR & $P_{R}$ (\%,

observation)
& 
$V_\text{rot}$
(Stevance's model) & 
$V_\text{rot}$
(Current model) &
$\log(j/1\,\text{cm}^2\,\text{s}^{-1})$
\\
\hline 
93b & 0.077 & 324 & 277 & 17.88
\\
102 & 0.057 & 234 & 444 & 17.85

\\
\hline 
\end{tabular}
\end{table*}


In hydrodynamic simulations, we have considered an isothermal wind and assumed that variations of the effective temperature all over the stellar surface did not contribute to the polarization.
In order to see the effect due to the change in the temperature, we computed the intrinsic polarization from  Eq.~\eqref{tempol}, where the temperature changes due to the effective gravity (gravity darkening) and the number density is assumed to be radial dependent ($n_\text{e}(r)\propto \dot{M}/r^2v(r)$). For WR93b star with rotation $\omega=20 \%$ of the critical velocity, the obtained polarization was found to be $0.0027\%$, which is smaller than the upper limit obtained from the observation \citep{2018MNRAS.479.4535S}, and this means that this effect is not sufficient to break the symmetry of the star.
On the other hand, for the critical rotation, the polarization can reach up to $14.52\%$. For WR102 with rotation $\omega=20\%$, the polarization is similar to the previous star and has a value of $0.006\%$, 
and the rotation at critical velocity leads 
 to a polarization with the value of $32.5\%$.

\section{Discussion}\label{discus}

We have determined the polarization due to  electron scattering from hydrodynamical models of optically thin winds. The analysis involves several simplifications, including the neglect of the optically thick part of the wind on the density distribution and polarization which are perhaps the most significant ones. Therefore, in this section, we discuss the effect of the optical depth on the density distribution and polarization, as WR stars are known to have optical thick winds.

\subsection{Density distribution}

The adopted density distribution was derived from hydrodynamical models appropriate for winds which are optically thin in continuum. However, winds of WR stars may be optically thick and the radius of the hydrostatic core could be significantly lower than the radius of the unity mean optical depth \citep{2012A&A...538A..40G,2021A&A...647A.151P}.
If we consider the effect of gravity darkening, the density has a prolate distribution;  however, in the optically thick regime, the radiative diffusion effects may further complicate the structure of the flow \citep{2004IAUS..215..527G}.
Using the model outlined in
\citet{1998Ap&SS.260..149O}, and \citet{2002ApJ...581.1337D}, which was adopted to optically thick winds, the mass flux is expressed as
\begin{equation}\label{massflux}
   \dot{m}(\theta) =\dot{m}(\theta=0)(1-\omega^2\sin^2\theta),
\end{equation}

where $\dot{m}(0)$ is the mass flux at the pole.
The mass conservation implies that in addition to the mass flux, the wind speed would be modified by the rotational effect.
The
wind terminal speed is expressed as a function of colatitude,
\begin{equation}
   \varv_\infty (\theta)= \varv_\infty(\theta=0)\sqrt{1-\omega^2\sin^2\theta},
\end{equation}
where $\varv_\infty(0)$ is the terminal speed at the pole.
For $\omega=20\%$, the polarization is $-0.48\%$ and $-1.03\%$ for WR93b and for WR102, respectively;  also, at the critical rotation, the polarization 
is $-20.5\%$ for WR93b and $-43\%$ for WR102. Comparing this with the hydrodynamic model including nonradial forces and gravity darkening, the polarization values showed comparable results, where, for example, $\omega=30\%$, $P_R =-1.1\%$, and $P_R= -2.3\%$ and from hydrodynamic with  nonradial forces and gravity darkening $P_R=-0.95$  and $P_R=-0.11$ for WR93b and for WR102, respectively.
We note that in the hydrodynamic model, we integrated from $r=1.2R_\star$ as mentioned earlier.
This indicates that the effect of modifying the wind density due to the optical thickness of WR star winds does not significantly change the main results of our paper.
\subsection{Optical depth}
The polarization does not only depend on the inclination or rotational velocity, but it also depends on the opacity of the medium.
In our model, we assumed the optical thin regime, but
using Monte Carlo radiative transfer, \citet{2019MNRAS.489.2873C} show that the polarization increases by increasing the optical depth ($\tau>1$). Due to their high mass loss rate, WR stars have an optical thick wind even to the electron scattering. To investigate the effect of the optical depth, using a more detailed treatment of the polarized radiative transfer, for instance with the Monte Carlo method of the radiative transfer, is needed.
A work is under way using the latter method in couple with the hydrodynamic model, but we note that our models account for the strongest effect which is due to the last scattering in an optically thin envelope.

We assumed that the stars produce unpolarized light, which propagates into the wind where each photon scatters once at most. This is a simplification, especially for the winds of WR stars with a high density. However, multiple scattering leads to depolarization; therefore, we selected integration radii outside the optically thick envelope.

\subsection{Magnetic field effect and the ionization structure}
In this  work, we   did not consider the magnetic field. As a result of the magnetic field, the wind flows along the field lines, the mass-flux depends on the magnetic field tilt, and the circumstellar environment deviates from spherical symmetry \citep{2008MNRAS.385...97U,2017AN....338..868K}. This should also result in intrinsic polarization.
Also, we neglected the variations in electron density due to the change of ionization, which is a reasonable approximation when the surface abundances are dominated by helium and for relatively high effective temperatures.

To maintain the acceleration around the sonic point, high effective temperatures are needed \citep{refId0},which allow for strong dependence of the iron peak element opacity on the temperature in WR stars. As showed by \citet{refId0}, the "Iron Bump" opacity may lead to an optically thick wind. Another factor that could impact the wind optical depth is the combination between the Eddington limit and the limit of rotation. As discussed by \citet{2000A&A...361..159M}, and \citet{2021A&A...647A..13G} and known as the $\Gamma \Omega$ limit, for high $\Gamma$, just moderate rotation is sufficient to increase the mass loss rates and as consequence the wind becomes optically thick. These effects can further modify our results.

\section{Conclusions}
In this work, we used the outputs of the density distribution from the hydrodynamic model to calculate the intrinsic polarization for WR93 and WR102 for different rotational velocities.
Oblate density distribution  leads to a positive polarization, while prolate distribution leads to a negative polarization, and this agrees with the model predicted by \citet{1977A&A....57..141B}.
The intrinsic polarization strongly depends on the rotational velocities and the %
inclination.
Based on our assumption, where we have scaled the density to take  realistic values of mass-loss rates of studied stars into account,
we show that fast rotation leads to a prolate density distribution around WR stars. This produces just a weak polarization signature, which is smaller than the upper limit of polarization from observations for low rotational velocities.
Therefore,  the limits of rotational velocities of WR stars from polarization do not contradict models of a long duration gamma-ray burst, according to which the fast rotating WR stars are progenitors of gamma ray bursts.

\begin{acknowledgements}

The authors thank Profs.~R.~Ignace
and S.~Owocki
for discussion.
The access to computing and storage facilities owned by parties and
projects contributing to the National Grid Infrastructure MetaCentrum, provided
under the program “Projects of Large Infrastructure for Research, Development,
and Innovations” (LM2010005) is appreciated. 
This research was supported by 
the grant Primus/SCI/17.
\end{acknowledgements}

%
   \bibliographystyle{aa} 
   \bibliography{references} 

\begin{thebibliography}{48}
\expandafter\ifx\csname natexlab\endcsname\relax\def\natexlab#1{#1}\fi

\bibitem[{{Abbott}(1982)}]{1982ApJ...259..282A}
{Abbott}, D.~C. 1982, \apj, 259, 282

\bibitem[{{Bjorkman} \& {Cassinelli}(1993)}]{1993ApJ...409..429B}
{Bjorkman}, J.~E. \& {Cassinelli}, J.~P. 1993, \apj, 409, 429

\bibitem[{{Blondin} {et~al.}(1990){Blondin}, {Kallman}, {Fryxell}, \&
  {Taam}}]{1990ApJ...356..591B}
{Blondin}, J.~M., {Kallman}, T.~R., {Fryxell}, B.~A., \& {Taam}, R.~E. 1990,
  \apj, 356, 591

\bibitem[{{Brown} {et~al.}(2000){Brown}, {Ignace}, \&
  {Cassinelli}}]{2000A&A...356..619B}
{Brown}, J.~C., {Ignace}, R., \& {Cassinelli}, J.~P. 2000, \aap, 356, 619

\bibitem[{{Brown} \& {McLean}(1977)}]{1977A&A....57..141B}
{Brown}, J.~C. \& {McLean}, I.~S. 1977, \aap, 57, 141

\bibitem[{{Brown} {et~al.}(1978){Brown}, {McLean}, \&
  {Emslie}}]{1978A&A....68..415B}
{Brown}, J.~C., {McLean}, I.~S., \& {Emslie}, A.~G. 1978, \aap, 68, 415

\bibitem[{{Carlos-Leblanc} {et~al.}(2019){Carlos-Leblanc}, {St-Louis},
  {Bjorkman}, \& {Ignace}}]{2019MNRAS.489.2873C}
{Carlos-Leblanc}, D., {St-Louis}, N., {Bjorkman}, J.~E., \& {Ignace}, R. 2019,
  \mnras, 489, 2873

\bibitem[{{Cassinelli} {et~al.}(1987){Cassinelli}, {Nordsieck}, \&
  {Murison}}]{1987ApJ...317..290C}
{Cassinelli}, J.~P., {Nordsieck}, K.~H., \& {Murison}, M.~A. 1987, \apj, 317,
  290

\bibitem[{{Castor} {et~al.}(1975){Castor}, {Abbott}, \&
  {Klein}}]{1975ApJ...195..157C}
{Castor}, J.~I., {Abbott}, D.~C., \& {Klein}, R.~I. 1975, \apj, 195, 157

\bibitem[{{Castor}(1974)}]{1974MNRAS.169..279C}
{Castor}, J.~L. 1974, \mnras, 169, 279

\bibitem[{{Chandrasekhar}(1950)}]{1950ratr.book.....C}
{Chandrasekhar}, S. 1950, {Radiative transfer.}

\bibitem[{{Chiosi} {et~al.}(1979){Chiosi}, {Nasi}, \&
  {Bertelli}}]{1979A&A....74...62C}
{Chiosi}, C., {Nasi}, E., \& {Bertelli}, G. 1979, \aap, 74, 62

\bibitem[{{Colella} \& {Woodward}(1984)}]{1984JCoPh..54..174C}
{Colella}, P. \& {Woodward}, P.~R. 1984, Journal of Computational Physics, 54,
  174

\bibitem[{{Collins}(1963)}]{1963ApJ...138.1134C}
{Collins}, George~W., I. 1963, \apj, 138, 1134

\bibitem[{{Collins} \& {Harrington}(1966)}]{1966ApJ...146..152C}
{Collins}, George~W., I. \& {Harrington}, J.~P. 1966, \apj, 146, 152

\bibitem[{{Conti}(1975)}]{1975MSRSL...9..193C}
{Conti}, P.~S. 1975, Memoires of the Societe Royale des Sciences de Liege, 9,
  193

\bibitem[{{Cranmer} \& {Owocki}(1995)}]{1995ApJ...440..308C}
{Cranmer}, S.~R. \& {Owocki}, S.~P. 1995, \apj, 440, 308

\bibitem[{{Dwarkadas} \& {Owocki}(2002)}]{2002ApJ...581.1337D}
{Dwarkadas}, V.~V. \& {Owocki}, S.~P. 2002, \apj, 581, 1337

\bibitem[{{Gayley}(2004)}]{2004IAUS..215..527G}
{Gayley}, K.~G. 2004, in Stellar Rotation, ed. A.~{Maeder} \& P.~{Eenens}, Vol.
  215, 527

\bibitem[{{Gr{\"a}fener}(2021)}]{2021A&A...647A..13G}
{Gr{\"a}fener}, G. 2021, \aap, 647, A13

\bibitem[{{Gr\"afener} \& {Hamann}(2005)}]{refId0}
{Gr\"afener}, G. \& {Hamann}, W.-R. 2005, A\&A, 432, 633

\bibitem[{{Gr{\"a}fener} {et~al.}(2012{\natexlab{a}}){Gr{\"a}fener}, {Owocki},
  \& {Vink}}]{2012A&A...538A..40G}
{Gr{\"a}fener}, G., {Owocki}, S.~P., \& {Vink}, J.~S. 2012{\natexlab{a}}, \aap,
  538, A40

\bibitem[{{Gr{\"a}fener} {et~al.}(2012{\natexlab{b}}){Gr{\"a}fener}, {Vink},
  {Harries}, \& {Langer}}]{2012A&A...547A..83G}
{Gr{\"a}fener}, G., {Vink}, J.~S., {Harries}, T.~J., \& {Langer}, N.
  2012{\natexlab{b}}, \aap, 547, A83

\bibitem[{{Harrington} \& {Collins}(1968)}]{1968ApJ...151.1051H}
{Harrington}, J.~P. \& {Collins}, George~W., I. 1968, \apj, 151, 1051

\bibitem[{Harris {et~al.}(2020)Harris, Millman, van~der Walt, Gommers,
  Virtanen, Cournapeau, Wieser, Taylor, Berg, Smith, Kern, Picus, Hoyer, van
  Kerkwijk, Brett, Haldane, del R{'{\i}}o, Wiebe, Peterson,
  G{'{e}}rard-Marchant, Sheppard, Reddy, Weckesser, Abbasi, Gohlke, \&
  Oliphant}]{harris2020array}
Harris, C.~R., Millman, K.~J., van~der Walt, S.~J., {et~al.} 2020, Nature, 585,
  357

\bibitem[{Hunter(2007)}]{Hunter:2007}
Hunter, J.~D. 2007, Computing in Science \& Engineering, 9, 90

\bibitem[{{Ignace} {et~al.}(1996){Ignace}, {Cassinelli}, \&
  {Bjorkman}}]{1996ApJ...459..671I}
{Ignace}, R., {Cassinelli}, J.~P., \& {Bjorkman}, J.~E. 1996, \apj, 459, 671

\bibitem[{{K{\"u}ker}(2017)}]{2017AN....338..868K}
{K{\"u}ker}, M. 2017, Astronomische Nachrichten, 338, 868

\bibitem[{{Kurf{\"u}rst} {et~al.}(2020){Kurf{\"u}rst}, {Pejcha}, \&
  {Krti{\v{c}}ka}}]{2020A&A...642A.214K}
{Kurf{\"u}rst}, P., {Pejcha}, O., \& {Krti{\v{c}}ka}, J. 2020, \aap, 642, A214

\bibitem[{{MacFadyen} \& {Woosley}(1999)}]{1999ApJ...524..262M}
{MacFadyen}, A.~I. \& {Woosley}, S.~E. 1999, \apj, 524, 262

\bibitem[{{Maeder} \& {Meynet}(2000)}]{2000A&A...361..159M}
{Maeder}, A. \& {Meynet}, G. 2000, \aap, 361, 159

\bibitem[{McKinney(2011)}]{mckinney2011pandas}
McKinney, W. 2011, Python for High Performance and Scientific Computing, 14

\bibitem[{{McLean} {et~al.}(1979){McLean}, {Coyne}, {Frecker}, \&
  {Serkowski}}]{1979ApJ...231L.141M}
{McLean}, I.~S., {Coyne}, G.~V., {Frecker}, S.~J.~J.~E., \& {Serkowski}, K.
  1979, \apjl, 231, L141

\bibitem[{{Owocki} {et~al.}(1994){Owocki}, {Cranmer}, \&
  {Blondin}}]{1994ApJ...424..887O}
{Owocki}, S.~P., {Cranmer}, S.~R., \& {Blondin}, J.~M. 1994, \apj, 424, 887

\bibitem[{{Owocki} {et~al.}(1996){Owocki}, {Cranmer}, \&
  {Gayley}}]{1996ApJ...472L.115O}
{Owocki}, S.~P., {Cranmer}, S.~R., \& {Gayley}, K.~G. 1996, \apjl, 472, L115

\bibitem[{{Owocki} {et~al.}(1998){Owocki}, {Cranmer}, \&
  {Gayley}}]{1998Ap&SS.260..149O}
{Owocki}, S.~P., {Cranmer}, S.~R., \& {Gayley}, K.~G. 1998, \apss, 260, 149

\bibitem[{{Poniatowski} {et~al.}(2021){Poniatowski}, {Sundqvist}, {Kee},
  {Owocki}, {Marchant}, {Decin}, {de Koter}, {Mahy}, \&
  {Sana}}]{2021A&A...647A.151P}
{Poniatowski}, L.~G., {Sundqvist}, J.~O., {Kee}, N.~D., {et~al.} 2021, \aap,
  647, A151

\bibitem[{{Sander} {et~al.}(2012){Sander}, {Hamann}, \&
  {Todt}}]{2012A&A...540A.144S}
{Sander}, A., {Hamann}, W.~R., \& {Todt}, H. 2012, \aap, 540, A144

\bibitem[{{Schulte-Ladbeck}(1994)}]{1994Ap&SS.221..347S}
{Schulte-Ladbeck}, R.~E. 1994, \apss, 221, 347

\bibitem[{{Serkowski}(1970)}]{1970ApJ...160.1083S}
{Serkowski}, K. 1970, \apj, 160, 1083

\bibitem[{{Sobolev}(1960)}]{1960mes..book.....S}
{Sobolev}, V.~V. 1960, {Moving envelopes of stars}

\bibitem[{{Stevance} {et~al.}(2018){Stevance}, {Ignace}, {Crowther}, {Maund},
  {Davies}, \& {Rate}}]{2018MNRAS.479.4535S}
{Stevance}, H.~F., {Ignace}, R., {Crowther}, P.~A., {et~al.} 2018, \mnras, 479,
  4535

\bibitem[{{Tramper} {et~al.}(2015){Tramper}, {Straal}, {Sanyal}, {Sana}, {de
  Koter}, {Gr{\"a}fener}, {Langer}, {Vink}, {de Mink}, \&
  {Kaper}}]{2015A&A...581A.110T}
{Tramper}, F., {Straal}, S.~M., {Sanyal}, D., {et~al.} 2015, \aap, 581, A110

\bibitem[{{Ud-Doula} {et~al.}(2008){Ud-Doula}, {Owocki}, \&
  {Townsend}}]{2008MNRAS.385...97U}
{Ud-Doula}, A., {Owocki}, S.~P., \& {Townsend}, R. H.~D. 2008, \mnras, 385, 97

\bibitem[{{Vink} \& {de Koter}(2005)}]{2005A&A...442..587V}
{Vink}, J.~S. \& {de Koter}, A. 2005, \aap, 442, 587

\bibitem[{{Wood} {et~al.}(1993){Wood}, {Brown}, \& {Fox}}]{1993A&A...271..492W}
{Wood}, K., {Brown}, J.~C., \& {Fox}, G.~K. 1993, \aap, 271, 492

\bibitem[{{Woosley}(1993)}]{1993ApJ...405..273W}
{Woosley}, S.~E. 1993, \apj, 405, 273

\bibitem[{{Woosley} {et~al.}(1999){Woosley}, {Eastman}, \&
  {Schmidt}}]{1999ApJ...516..788W}
{Woosley}, S.~E., {Eastman}, R.~G., \& {Schmidt}, B.~P. 1999, \apj, 516, 788

\end{thebibliography}
%
\end{document}